\documentclass{iaus}
\usepackage{graphicx}
\usepackage{natbib}
                                                                                                                           
\newcommand{\cdr}{\ensuremath{^{13}\mem{C}}}
\newcommand{\sprn}{\mbox{$s$ process}}

\newcommand{\teff}{\ensuremath{T_{\rm eff}}}
\newcommand{\abb}[1]{Fig.\,\ref{#1}}

\newcommand{\kap}[1]{Sect.\,\ref{#1}}

\newcommand{\mem}[1]{\ensuremath{\mathrm{ #1}}}

\newcommand{\msun}{\ensuremath{\, {\rm M}_\odot}}

\title[CSPN evolution]{The Evolution of Central Stars of Planetary Nebulae}

\author[Herwig, Freytag \& Werner]   
{Falk Herwig$^1$, Bernd Freytag$^{1,2}$ and Klaus Werner$^3$} 

\affiliation{$^1$Theoretical Astrophysics Group, Los Alamos National Laboratory, Los Alamos, NM 87545, USA \break email: fherwig@lanl.gov \\[\affilskip]
$^2$ Department of Astronomy
and Space Physics at Uppsala University, Sweden  \break
email: bf@astro.uu.se \\[\affilskip]
$^3$ Institut f\"ur Astronomie und Astrophysik,
Universit\"at T\"ubingen, Sand~1, D-72076 T\"ubingen, Germany \break
email: werner@astro.uni-tuebingen.de
}

\pubyear{2006}
\volume{234}  
\pagerange{119--126}
\date{Draft: \today}
\jname{Planetary Nebulae in our Galaxy and Beyond}
\editors{M. J. Barlow \& R. H. M\'endez, eds.}
\begin{document}

\maketitle

\begin{abstract}
  The evolution of central stars of planetary nebulae can proceed in
  several distinct ways, either leading to H-deficiency or to H-normal
  surface composition. Several new simulations of the evolution
  channels that lead to H-deficiency are now available, mainly the
  born-again scenarios that are triggered by a He-shell flash during
  the hot pre-white dwarf evolution phase. A realistic AGB progenitor
  evolution is important for correct HRD tracks, that allow mass
  determinations. New hydrodynamic simulations of He-shell flash
  convection including cases with H-ingestion are now performed, and
  allow a determination of the convective extra-mixing efficiency.
  This has direct consequences for the intershell abundance
  distribution of AGB stars that can be observed in the H-deficient
  CSPN.  
\end{abstract}

\firstsection 

\section{Introduction}
The evolution of the central stars of planetary nebulae (CSPN) is
closely connected to the evolution of the progenitor phase, the
Asymptotic Giant Branch (AGB) stars, and has observable effects for
the CSPN progeny, the WD stars. The AGB evolution phase is the result
of previous main-sequence and horizontal-branch evolution of low- and
intermediate mass stars. An updated classification of mass ranges
involving AGB stars has been recently suggested by \citet[][Fig.\,
2]{herwig:04c}. Accordingly, low-mass stars are those below an initial
mass of $\approx 1.8\msun$ that ignite a He-core flash at the tip of
the RGB, while intermediate mass stars are those initially more
massive than $\approx 1.8\msun$ but less massive than required to
explode as a core-collapse supernova. The dividing mass at $\approx
10\msun$ separates those super-AGB stars that form ONe-white dwarfs
and those that ignite as core collapse supernova
\citep[e.g.][]{garcia-berro:94}. However, quantitatively the fraction
of AGB stars that explode as core-collapse supernova is still very
uncertain, largely due to uncertainties in super-AGB mass loss, and
the physics of mixing and burning, e.g. during dredge-up and
hot-bottom burning (see below). For AGB stars it is useful to use the
limiting mass for hot-bottom burning (HBB) as a dividing line for
low-mass and massive AGB stars. Hot-bottom burning transforms dredged
up C into N and prevents the formation of C-rich composition
\citep[e.g.][]{scalo:75,lattanzio:92b}. HBB has therefore an important
influence on the evolution of AGB star, and the composition of the
ejecta that later form the planetary nebula. AGB stars that are
massive enough to ignite carbon burning are the super-AGB stars.

Conceptually, the evolution of AGB stars is well understood, at least
at solar and modest metal deficiency, like in the Magellanic Clouds.
The review by \citet{iben:83b} describes the basic properties of AGB
stars: thermal pulses, the necessity of the third dredge-up, the
effect of mass loss and the potential for important nucleosynthesis.
More recent reviews by \citet{lattanzio:97} and \citet{bloecker:99b}
include accounts of hot-bottom burning, and an improved quantitative
description of AGB evolution. Dedicated reviews were written on mass
loss \citep{wilson:00}, the \sprn\ \citep{busso:99,meyer:94} and the
cool post-AGB stars that allow important observations concerning the
\sprn\ \citep{vanwinckel:03}.

\begin{figure}
\begin{center}
 \includegraphics[width=10.5cm]{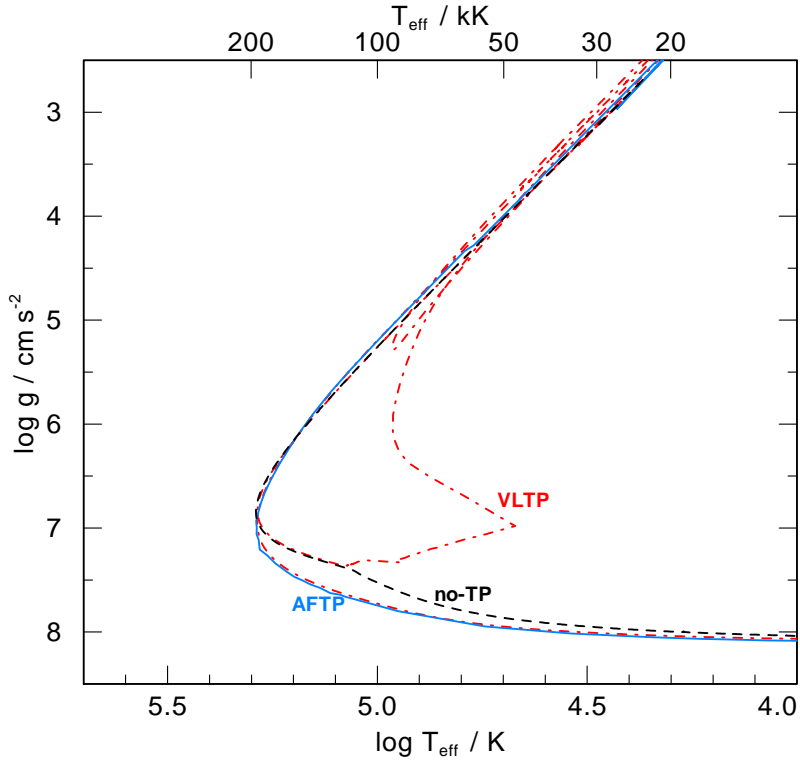}  
  \caption{ Comparison of three different evolution channels in the
$g$--\teff--plane for a $0.604\msun$ post-AGB star that had a
main-sequence mass of $2\msun$ and evolved through about a dozen
thermal pulses with dredge-up. This figure originally appeared in the
Publications of the Astronomical Society of the Pacific
\citep{werner:06}.  Copyright 2005, Astronomical Society of the
Pacific; reproduced with permission of the Editors.
}\label{fig:fig_gteff_herwig}
\end{center}
\end{figure}
Unfortunately, major unresolved issues remain, that severely limit the
predictive quality of current AGB modeling \citep{herwig:04c}. The
issues concern the mass-loss history, the nature and importance of
hydrodynamic mixing, including the physics of mixing at the bottom of
the envelope convection zone that seems to be critical for the third
dredge-up, and for formation of the assumed \cdr-pocket. Another
problem is the rather poorly known effect of rotation and magnetic
fields on the evolution, nucleosynthesis and formation of asymmetries
of bi-polar structure in proto-planetary nebulae. All these problems
are amplified for stars at very low metallicity, partly because less
observational checks are available.

The evolution of CSPN (described in \kap{sec:mod-cspn}) depends
sensitively on the progenitor AGB evolution. In particular, the
observed abundances of the H-deficient CSPN are linked to mixing
processes in the He-shell flash of the AGB phase, and some new results
in this area are presented in \kap{sec:hydro}.

\section{Evolution of CSPN}
\label{sec:mod-cspn}
We have now modern stellar evolution models of several distinct
channels of post-AGB evolution, and through the CSPN phase
\citep{werner:06}.  These channels are characterized by the thermal
pulse cycle phase at which they depart from the AGB. The most common is the
undisturbed evolution off the AGB, through the CSPN phase and onto the
cooling track of white dwarfs. These CSPN should show a surface
abundance pattern that reflects the former AGB envelope including
enrichment of that envelope through dredge-up \citep{napiwotzki:99b}.

The possibility of a born-again evolution was recognized early, by
\citet{schoenberner:83} and \citet{iben:83a}. Several more recent
calculations are now available \citep{herwig:99c,lawlor:03}, including
a new grid of H-deficient CSPN tracks based on the born-again scenario
by \citet{miller-bertolami:06}.  Two flavours of the born-again
evolution are now distinguished: in the very late thermal pulse (VLTP)
the remaining small amount of H in the surface layers is convectively
mixed into the He-shell flash convection, while in the late thermal
pulse (LTP) this does not happen \citep{bloecker:00a,herwig:00a}.  It
should be noted that the physics of convective-reactive mixing which
characterizes the VLTP is not well understood, as emphasized by
\citet{herwig:01a}.  Currently, efforts are underway at Los Alamos and
elsewhere to study the hydrodynamic aspects of this problem. The
simulations discussed in the following section are a first step into
that direction.

 According to their different nucleosynthetic and
mixing evolution the VLTP and LTP cases will produce CSPN with
observationally distinguishable properties. For example, while
Sakurai's object and V605 Aq are now considered the result of a VLTP
evolution, FG Sge is the result of a LTP evolution, and the arguments
in the literature leading to this conclusion have been reviewed
recently by \citet{werner:06}.

\begin{figure}
\begin{center}
 \includegraphics[width=10.5cm]{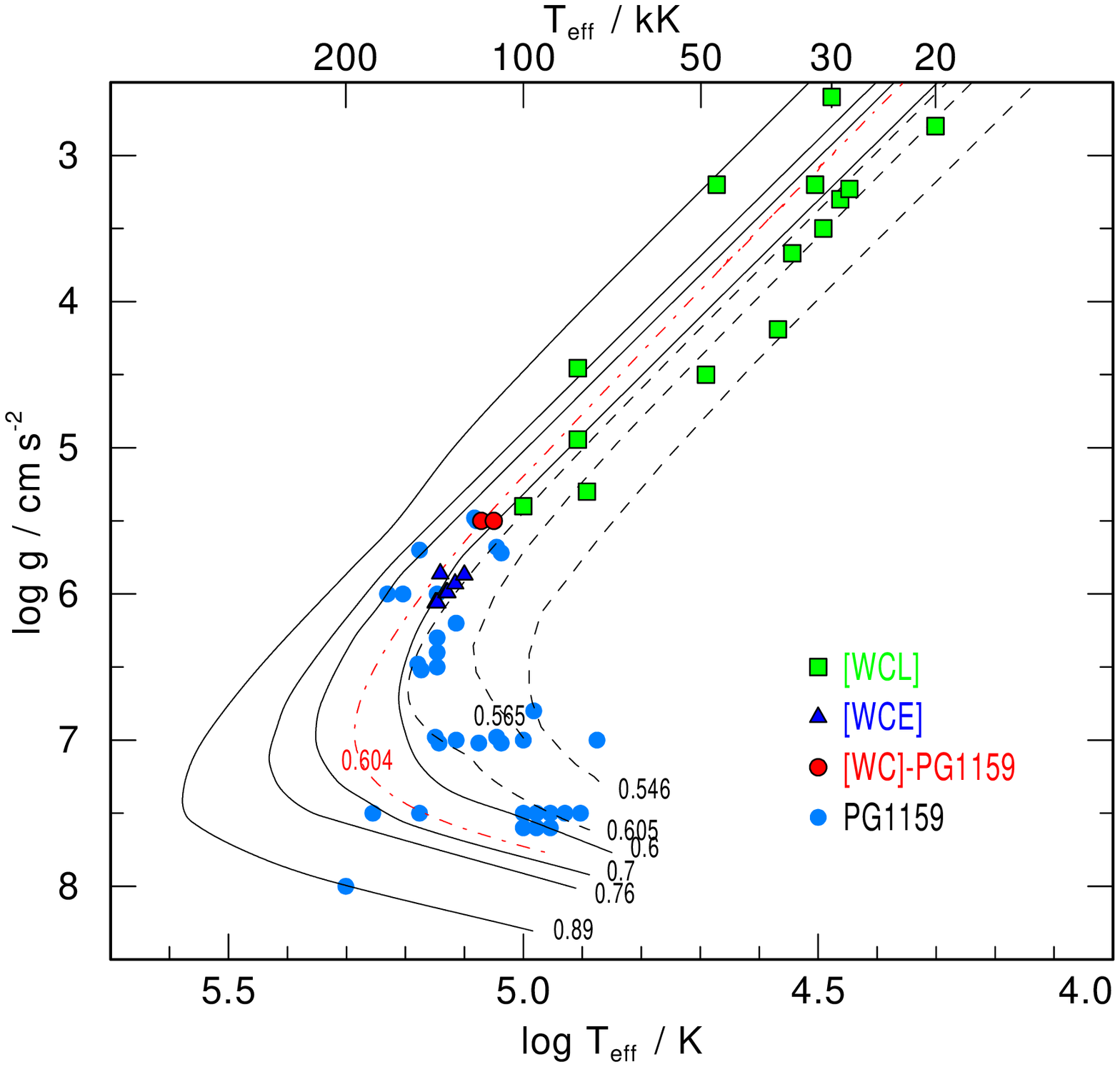} 
  \caption{ H-normal evolutionary tracks (labels: mass in M$_\odot$)
    are from \citet{schoenberner:83} and \citep{bloecker:95b} (dashed
    lines) and \citet{wood:86}. The $0.604\msun$ track
    \citet{2003IAUS..209..111H,werner:06} (dot-dashed line) is the
    final CSPN track following a VLTP evolution and therefore has
    H-deficient composition.  Shown as well are hot hydrogen-deficient
    post-AGB stars in $g$--\teff--plane. Wolf-Rayet central stars of
    early and late type ([WCE], [WCL], from \cite{hamann:97}), PG1159
    stars as well as two [WC]--PG1159 transition objects (Abell ~30
    and 78) \citep[see][for ref.]{werner:06} \teff\ for the [WC] stars
    is related to the stellar radius at $\tau_{\rm Ross}$=20. This
    figure originally appeared in the Publications of the Astronomical
    Society of the Pacific \citep{werner:06}.  Copyright 2005,
    Astronomical Society of the Pacific; reproduced with permission of
    the Editors.  }\label{fig:fig_gteff}
\end{center}
\end{figure}

In addition to the two born-again channels there may be a possibility
for a H-deficient, but not H-free CSPN evolution without the
born-again sequence of events. The AGB final thermal pulse (AFTP)
scenario requires some mechanism to shed the remaining small envelope
mass exactly after the final AGB thermal pulse with efficient
dredge-up. In view of the physics uncertainties, in particular of
envelope convection, this scenario is not inconceivable.
\citet{demarco:02b} and \citet{demarco:03a} have tentatively
investigated the spiraling-in and swallowing of a massive planet or
very low mass stellar companion as an enhanced mass-loss process for
the AFTP.  The AFTP scenario is currently the most likely scenario for
the hybrid CSPN \citep{napiwotzki:91}.

\abb{fig:fig_gteff_herwig} shows the comparison of different evolution
channels. These calculations give an estimate of the difference of the
tracks due to different surface chemistry and evolution history. In
particular the highest temperature is the same for all channels. 

More important for the location of the track in the HRD, and therefore
for mass determinations using stellar-evolution tracks, is the
progenitor evolution. \abb{fig:fig_gteff} shows CSPN from various
authors. The more recent 0.604\msun\ track is hotter than older tracks
of the same mass. This difference is largely due to differences in the
assumptions about AGB mass loss and the treatment of third dredge-up.
As already shown by \citet{bloecker:95b} CSPN with the same core mass
can have significantly different tracks in the HRD. A core descending
from an initially more massive progenitor simulated with a higher mass-loss 
rate will follow a hotter track than a core of the same mass with
a lower-mass progenitor with lower mass loss. Similarly, efficient
third dredge-up leads to a hotter CSPN track compared with a core
descending from a lower-mass progenitor with less third dredge-up.
Binarity may influence the relation between core mass and HRD CSPN
track additionally, through common envelope evolution, tidal
synchronization, enhanced mass loss etc.

\section{Hydrodynamic simulations of He-shell flash convection}
\label{sec:hydro}

\begin{figure}
 \includegraphics{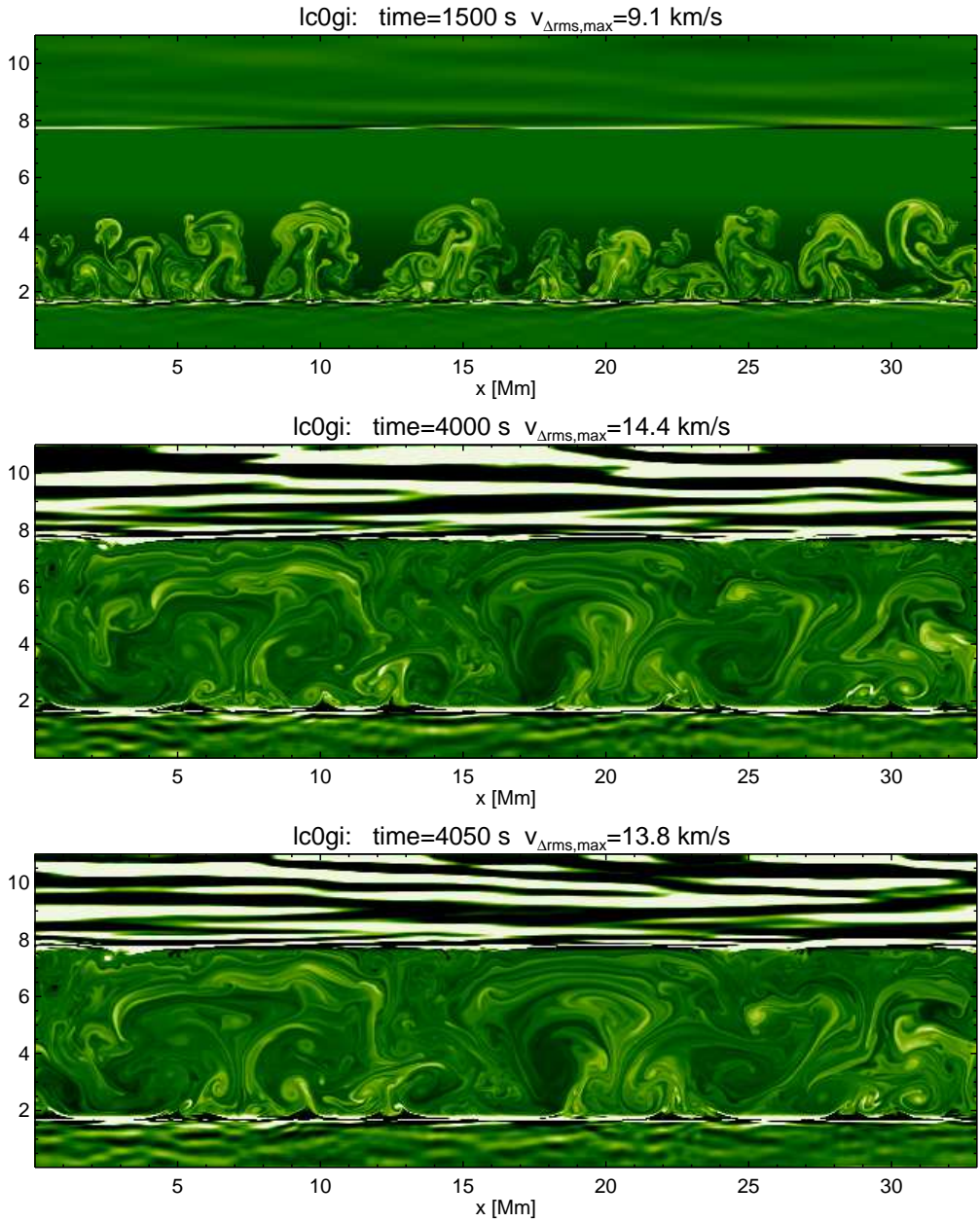} 
  \caption{Entropy inhomogeneities for three snapshots of a 2D simulation
  with a resolution of 2400$\times$800 grid points.
  The horizontal average of the entropy has been subtracted
  to render small fluctuations visible.
  A brighter color indicates material with an entropy excess (i.e.\ lower
  density than the surroundings).
  The overturning flow inside the convection zone clearly differs
  from the oscillatory motions due to gravity waves inside the stable layers.
  }\label{fig:flow}
\end{figure}

\begin{figure}
 \includegraphics{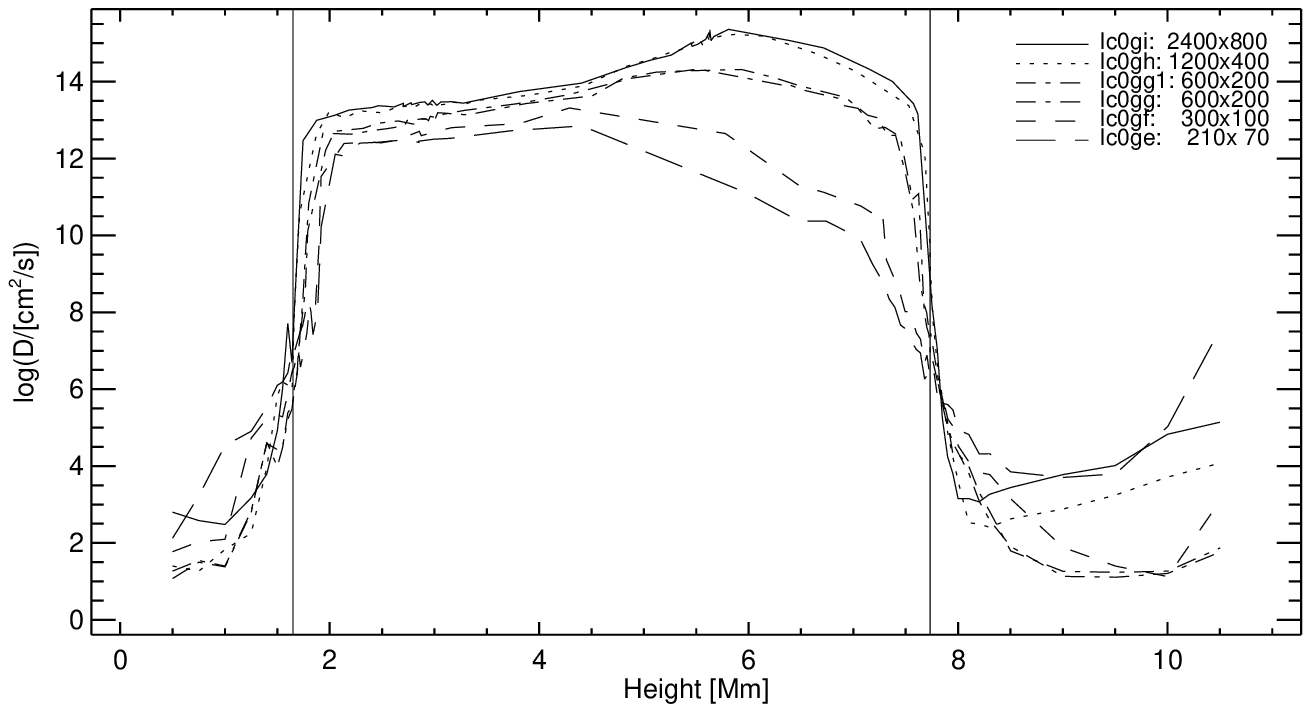} 
  \caption{Logarithm of the diffusion coefficient versus height
  for models with different resolutions (indicated in the legend).
  The two vertical lines mark the boundaries of the entropy plateau
  at the start of the simulations. Later, they are slightly smeared out.
  The results from the runs with the two lowest resolutions differ
  significantly from the runs with next higher resolutions.
  Nevertheless, the agreement between the two highest-resolution runs
  is rather good.
  }\label{fig:Ddiff}
\end{figure}

We used the RAGE code to perform hydrodynamics simulations
of a convection zone under conditions closely resembling those encountered
in a typical He-flash during the
hot pre-white dwarf evolution phase.
The initial stratification is in hydrostatic equilibrium and
consists of an entropy plateau,
where the convection zone is expected to develop,
with two adjacent stable regions.
The convection is driven by a heat source in a thin layer
at the bottom of the plateau to simulate the effect of He burning.
An number of 2D and 3D runs with various resolutions and heating rates
was produced.
See \citet{herwig:06a} for more details.

The first panel in Fig.~\ref{fig:flow} shows the spatial
entropy inhomogeneities at a snapshot during the onset of convection.
Hot plumes rise from the bottom of the entropy plateau.
They merge while they grow in amplitude and vertical extent
and start to excite gravity waves in the upper stable layers well
before actually reaching the upper boundary.
Later, there are typically four large instationary convective cells
with a number of smaller eddies
that continue to drive gravity waves in the adjacent layers
\citep{press:81}.

The transition between convectively unstable and stable layers
appears to be very sharp -- without any plumes penetrating
significantly into the the stable region.
This is very different from the situation of thin stellar surface
convection zones \citep{freytag:96}.
However, $k$-$\omega$-diagrams for the vertical velocity
at various height levels
clearly show the mode signature of the gravity waves
well inside the convection zone
and also evidence of the convective flow reaching into
the stable layers \citep{herwig:06a}.

Of particular interest are the mixing properties across the convection boundaries.
The mixing is efficient inside the convection zone.
Sometimes blobs of relative hot material are
sheared of the boundary to the top stable zone
(see e.g.\ the tiny bright spots near the top left of the convection
zone in the middle and bottom panels in Fig.~\ref{fig:flow})
and are transported
rapidly into deeper layers once they are entrained in a downflow region.
Future simulations that include various species and nuclear reactions
will show what happens if H-rich material is brought in this way
into layers where it can burn.

The velocities in the stable layers are not much smaller than in the
convection zone.  However, wave motions are almost reversible and do
not contribute much to the mixing.  To quantify the mixing the
trajectories of tracer particles (150 particles per horizontal layer
in 68 layers) have been integrated based on snapshots with a high
sampling rate of 0.5\,s.  The spread of the relative vertical
positions -- or alternatively the relative entropy -- of the particles
in each ensemble with time gives a measure of the diffusion
coefficient with height, plotted in Fig.~\ref{fig:Ddiff} for models
with different numerical resolutions.  The large diffusion coefficient
within the convection zone drops rapidly near the boundaries.  This
decay reaches slightly into the stable regions and can be described by
one (at the top) or two (at the bottom) exponentials
\citep{freytag:96}.

The determination of the very small mixing coefficients far away from
the convection zone is not very reliable. This small amount of mixing
in the stable layer is due to internal gravity waves, and may be far
reaching.

\section{Conclusions}
Among the CSPN channels, the evolution leading to H-deficient surface
composition is of great importance for constraining nucleosynthesis
and mixing in the AGB progenitors. A clear correlation exists, e.g.
between the mixing efficiency at the bottom of the He-shell flash
convection and the O abundance in the intershell that is subsequently
observed in H-deficient CSPN. AGB calculations with parameterized
overshoot calculations by \citet{herwig:99a} showed that more
convective extra-mixing leads to higher O abundance, and that an
exponential decay parameter for extra-mixing (see \citet{herwig:99a}
for the definition of) $f \approx 0.01$ would be consistent with CSPN
observed O abundances in the range $8 - 20\%$ by mass. Our preliminary
analysis of hydrodynamic simulations of the He-shell flash convection
zone quantitatively agrees with this level of convective extra-mixing.
We regard this agreement of hydrodynamic simulation and astrophysical
observation as a successful validation of the computational
simulations, and feel encouraged to continue this work towards more
complex simulations of convective-reactive mixing.

\acknowledgements 
This work was carried out in part under the auspices
of the National Nuclear Security Administration of the U.S. Department
of Energy at Los Alamos National Laboratory under Contract No.
DE-AC52-06NA25396, and funded by NNSA's Advanced Simulation and
Computing (ASC), Verification and Validation Program.

\end{document}